\documentclass[english,aps,preprint]{revtex4}
\usepackage{graphicx}

\begin{document}

\preprint{This line only printed with preprint option}

\title{Could humans recognize odor by phonon assisted tunneling?}

\author{Jennifer C. Brookes}

\email{j.brookes@ucl.ac.uk}

\affiliation{Department of Physics and Astronomy, University College London, Gower
Street, London WC1E 6BT, United Kingdom}

\author{Filio Hartoutsiou}

\email{to_milaraki@hotmail.com}

\affiliation{Department of Physics and Astronomy, University College London, Gower
Street, London WC1E 6BT, United Kingdom}

\author{A. P. Horsfield}

\email{a.horsfield@ucl.ac.uk}

\affiliation{Department of Physics and Astronomy, University College London, Gower
Street, London WC1E 6BT, United Kingdom}

\author{A. M. Stoneham}

\email{a.stoneham@ucl.ac.uk}

\affiliation{Department of Physics and Astronomy, University College London, Gower
Street, London WC1E 6BT, United Kingdom}

\begin{abstract}
Our sense of smell relies on sensitive, selective atomic-scale processes
that are initiated when a scent molecule meets specific receptors
in the nose. However, the physical mechanisms of detection are not
clear. While odorant shape and size are important, experiment indicates
these are insufficient. One novel proposal suggests inelastic electron
tunneling from a donor to an acceptor mediated by the odorant actuates
a receptor, and provides critical discrimination. We test the physical
viability of this mechanism using a simple but general model. Using
values of key parameters in line with those for other biomolecular
systems, we find the proposed mechanism is consistent both with the
underlying physics and with observed features of smell, provided the
receptor has certain general properties. This mechanism suggests a
distinct paradigm for selective molecular interactions at receptors
(the swipe card model): recognition and actuation involve size and
shape, but also exploit other processes.
\end{abstract}
\maketitle
Our sense of smell affects our behavior profoundly. Discrimination
between small molecules, often in very low concentrations, allows
us to make judgments about our immediate environment\cite{leffingwell-2001-a}
and influence our perceptions. Even though odorants are key components
of many commercial products\cite{turin-2005-a}, the biomolecular
processes of olfaction are inadequately understood: scent design is
not straightforward. We know that odor detection involves several
types of receptor for a given odorant, and understand how a receptor
signal is amplified and processed\cite{axel-2005-a,buck-2005-a}.
However, the initial selective atomic-scale processes as the scent
molecule meets its nasal receptors are not well understood. Odorant
shape and size are certainly important, but experiment shows these
are insufficient. Here we assess the novel proposal that a critical
early step involves inelastic electron tunneling mediated by the odorant.
We test the physical viability of this mechanism\cite{bialek-1987-a}
using electron transfer (ET) theory, with values of key parameters
in line with those for other biomolecular systems. The proposed mechanism
is viable (there are no physics-based objections and is consistent
with known features of olfaction) provided the receptor has certain
general properties. This mechanism has wider importance because it
introduces a distinct paradigm for selective actuation of receptors:
whereas lock and key models\cite{silverman-2002-a} imply size, shape
and non-bonding interactions (the docking criteria) are all, in our
swipe card model recognition and actuation involve other processes
in addition to docking. Thus it encompasses and goes beyond mechanisms
such as proton transfer, discussed by us previously\cite{wallace-1993-a}.

All current theories agree that selective docking of odorants is important\cite{turin-2005-a}.
However, odorants are small molecules (rarely more than a few tens
of atoms\cite{turin-2005-a}), and it is improbable that docking criteria
alone offer sufficient discrimination. For example, molecules with
almost identical shapes can smell very different: replacing carbon
with its isosteres Si, Ge and Sn invariably markedly alter odor character\cite{wrobel-1982-b}.
Something more is needed for olfaction, leading to early suggestions
that odorant vibration frequencies were critical \cite{dyson-1938-a,wright-1982-a},
though without specific mechanisms. Both infrared and inelastic electron
tunneling\cite{lambe-1968-a,adkins-1985-a} (IETS) spectroscopies
distinguish very precisely between different molecules through vibrational
frequencies and intensities, which makes the proposal appealing. The
first specific mechanism (based on IETS) was Turin's\cite{turin-1996-a}
idea that there is odorant mediated inelastic tunneling of an electron
at the receptor: inelastic tunneling between receptor electronic states
differing in energy by $\hbar\omega$ that occurs only when energy
is conserved by emission of an odorant phonon of the right energy,
hence selectivity. Clearly, the vibration must also couple to the
electronic transition.

Experiment offers several tests of Turin's idea. It explains why certain
molecules with very different shapes can smell similar (e.g. boranes
and thiols), but also why some molecules of essentially identical
shape smell utterly different (e.g. 1,1-dimethylcyclohexane and its
sila counterpart) because of frequency or coupling changes. The question
of whether humans can distinguish between a molecule and its deuterated
counterpart is still controversial. There is evidence both for\cite{turin-1996-a,haffenden-2001-a}
and against\cite{keller-2004-a}. In animals, isotope discrimination
is well-documented\cite{havens-1995-a}. Both left- and right-handed
forms of enantiomers should have the same vibrational spectrum. The
odors of some enantiomer pairs are the same (type 1) while others
differ (type 2)\cite{brenna-2003-a}. Type 2 can be explained by docking
criteria (different chiralities fit different receptors), while type
1 is naturally explained by vibrational frequency. However, docking
and frequency together can account for both since chirality will affect
the intensity of response of receptors to the enantiomers (the helices
that form the walls of the receptors are chiral). Quantitative support
for the theory comes from the successful correlation of odor character
with tunneling frequency spectrum\cite{turin-2002-a} for a range
of odorants. Indeed, vibrational frequency has been found to correlate
\emph{better} with odor than structure \cite{takane-2004-a}. Thus
a molecule's vibrational spectrum appears closely linked to its odor.
We now test whether the physical processes underlying Turin's proposed
mechanism for detecting the frequency are credible.

First we focus on the odorant and perform a simple test of whether
odor can be related to vibrational frequency and to coupling to the
odorant charges (as required by IETS). We computed the vibrational
spectra of $H_{2}S$ and four boranes (decaborane, m-, o- and p-carborane).
The boranes are structurally similar, but all quite distinct from
$H_{2}S$. However, $H_{2}S$ and decaborane smell sulfuraceous, while
the carboranes smell camphoraceous. Using Gaussian 03\cite{g03} we
computed vibrational frequencies and infrared (IR) couplings, defined
as $\left|\partial\vec{p}/\partial Q_{i}\right|^{2}$ with $\vec{p}$
the dipole moment and $Q_{i}$ a displacement along normal mode $i$.
The sulfuraceous smell of $H_{2}S$ is associated with vibrations
in the region of $2600\,{\rm cm}^{-1}$\cite{crc2005}. In this region
decaborane has IR couplings that are one to two orders of magnitude
greater than the carboranes. Assuming that the IR couplings are a
good estimate of the electron-oscillator coupling in an olfactory
receptor, this could explain sulfuraceous and less sulfuraceous odors.

\begin{figure}[h]
\begin{centering}\includegraphics[width=0.9\columnwidth,keepaspectratio]{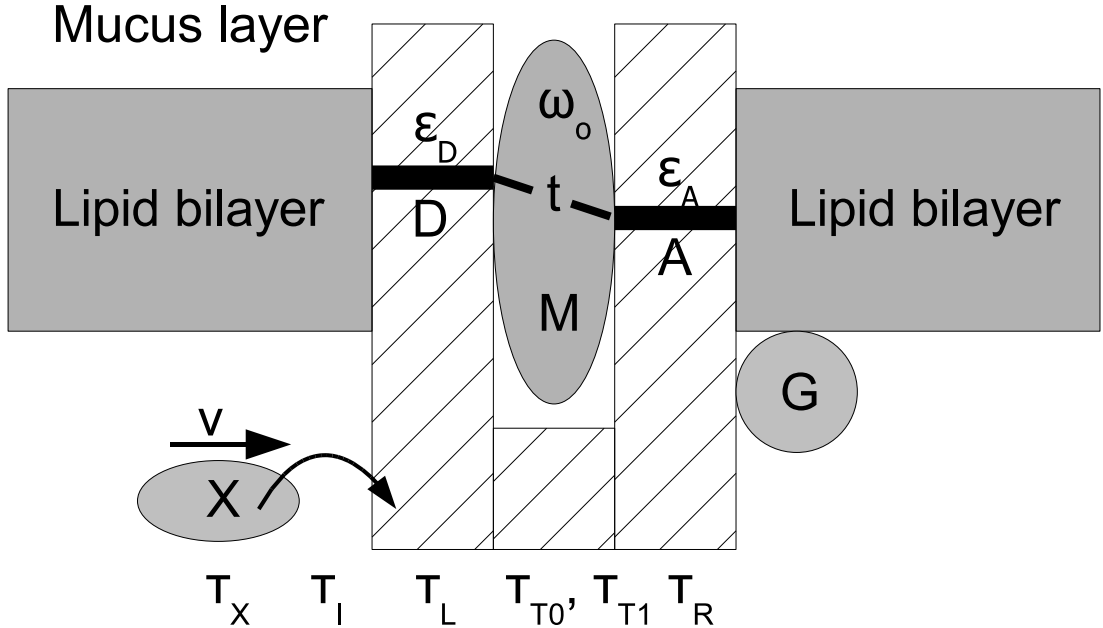}\par\end{centering}

\caption{\label{cap:atom chain}The olfactory receptor is a G-protein coupled
receptor with seven hydrophobic helices that span the cell membrane.
It responds to the arrival of a recognized odorant by releasing the
$\alpha$-subunit of a neighboring G-protein, which in turn initiates
a large influx of Ca ions into the cell, a signal that can be communicated
to the brain. This figure represents the model of the receptor we
use to describe its action. The electron (hole) source $X$ is likely
to be a reducing (oxidizing) agent in the cell fluid. It arrives at
a site on the outside of the receptor protein where it can exchange
charge after an average interval $\tau_{X}$. Once in place, it exchanges
a charge with a characteristic time $\tau_{I}$. The charge then travels
to the donor $D$ in one transmembrane helix of the protein over an
average time of $\tau_{L}$, from where it then hops to the acceptor
$A$ (possibly in a different helix) with either an average time $\tau_{T0}$
for non-discriminating ({}``elastic'') tunneling or an average time
$\tau_{T1}$ for discriminating ({}``inelastic'') tunneling. Only
the inelastic contribution is sensitive to the odorant ($M$) oscillator
frequency $\omega_{0}$, and so needs to dominate the elastic contribution
($\tau_{T0}\gg\tau_{T1}$). The electron then travels from $A$ to
trigger the release of the G-protein ($G$) over a time $\tau_{R}$.
Note that the terms elastic and inelastic refer \emph{only} to energy
exchange with the odorant.}
\end{figure}

Turning to the combined odorant and receptor system, we must determine
whether ET is possible on the relevant time scale, and if so whether
the discriminating electron transfer rate (with excitation of the
critical odorant mode) is sufficiently large relative to rates for
non-discriminatory transfer channels (without excitation of the critical
mode). This is necessary because most IETS experiments observe inelastic
tunneling with phonon emission as a weak adjunct to the elastic component,
detected only by complex post-processing of a type unlikely in a nasal
environment. Too little is currently known about the atomic and electronic
structure of real odorant receptors for full-scale calculations. Instead
we make general assumptions about the nature of the receptor, the
odorant and their interaction. These assumptions relate to a series
of characteristic times corresponding to the required physical processes
(see Fig. \ref{cap:atom chain}).

Turin's theory requires a source of electrons or holes to allow charge
flow to take place. The precise biological origin is not known, but
may well consist of reducing (oxidizing) species ($X$) in the cell
fluid \cite{turin-2005-a}. These molecules diffuse through the aqueous
medium and arrive with an average interval of $\tau_{X}$. Using a
standard approach for computing reactant collision rates in solution
from the diffusion equation and the Stokes-Einstein relation for the
diffusion coefficient\cite{atkins-2002-a} we get $\tau_{X}=3\eta/2n_{X}k_{B}T$
where $\eta$ is the viscosity of water ($0.891\times10^{-3}\,{\rm kgm}^{-1}{\rm s}^{-1}$),
$n_{X}$ is the concentration of $X$, $k_{B}$ is Boltzmann's constant
and $T$ is the temperature. Note that this result is independent
of the nature of $X$ or the receptor. Since $n_{X}$ will probably
lie in the range $1\,\mu{\rm M}\to100\,\mu{\rm M}$,  we get a  range
of values for $\tau_{X}$ of $10\,\mu{\rm s}\to1\,{\rm ms}$. The
charge now has to cross from the molecule to the receptor molecule,
a process that can be described by Marcus theory \cite{marcus-1964-a,ulstrup-1979-a,ProteinElectronTransfer},
and characterized by a time $\tau_{I}$. In proteins times range from
about $1\,{\rm ms}$ down to about $1\,\mu{\rm s}$\cite{gray-1996-a}.
The injected charge has to propagate through to the donor ($D$).
The route is not known, but probably involves hopping transport. Thus
the journey time is likely be in the ms to $\mu{\rm s}$ range as
for charge injection. The next step is the inelastic tunneling from
$D$ to $A$ (the acceptor), and it is this charge movement that actuates
the receptor. Now the charge must reach the mechanism that releases
the G-protein which in turn initiates the signal that is sent to the
brain. Again, we do not know the route taken but is likely to involve
charge hopping. So the characteristic time $\tau_{R}$ will probably
be in the range ms to $\mu{\rm s}$. Thus, overall charge injection
and extraction together are likely to occur on typical biological
time scales of $\mu{\rm s}$ to ms.

For the mechanism to work there must be essentially no tunneling from
$D$ to $A$ in the absence of the odorant,  either  because the distance
is too great or energy conservation is problematic. The odorant must
make inelastic transmission possible by a mechanism coupling electron
movement from $D$ and $A$ to vibrational excitation in the odorant.
In IETS there is a strong contribution to this coupling from the Coulomb
interaction between partial charges associated with oscillating atoms
and a mobile electron\cite{kirtley-1976-a,kirtley-1979-a}. This same
mechanism allows us to account for observed features of olfaction
including the detection of oscillators buried inside a molecule (e.g.
2,6 di-t-butyl phenol \cite{turin-2005-a}), and is compatible with
our swipe card model: the long-ranged interaction can couple the mobile
electron to the oscillator even with a loose fit.

The times characterising elastic ($\tau_{T0}$) and inelastic ($\tau_{T1}$)
ET from $D$ to $A$ are central to the success or failure of Turin's
mechanism. We treat $D$ and $A$ as single molecular orbitals with
energies $\varepsilon_{D}$ and $\varepsilon_{A}$, coupled to each
other by a weak hopping integral $t$, but not coupled to other electronic
states. Since the hopping between $D$ and $A$ is slow on electronic
time scales, the remaining electronic couplings must be very weak
to prevent electron leakage. However, $D$ and $A$ will be coupled
to oscillators in the odorant, receptor protein and the wider environment.
The ET rate from $D$ to $A$ can be computed from standard theory\cite{marcus-1956-a,marcus-1964-a,marcus-1965-a,ulstrup-1979-a,song-1993-a,ProteinElectronTransfer}
but with the odorant oscillator treated explicitly. We consider one
odorant oscillator of frequency $\omega_{o}$ which couples with strength
$\gamma_{D}$ ($\gamma_{A}$) to $D$ ($A$) . The environment is
treated as many oscillators with frequencies $\omega_{q}$ and coupling
strengths $\gamma_{qD}$ and $\gamma_{qA}$. The complete system is
described by the Hamiltonian $\hat{H}=\hat{H}_{D}+\hat{H}_{A}+\hat{v}$,
where $\hat{H}_{X}=|X\rangle\langle X|\left(\varepsilon_{X}+\hat{H}_{osc}+\hat{H}_{e-osc,X}\right)$
($X$ is $D$ or $A$), $\hat{v}=t\left(|D\rangle\langle A|+|A\rangle\langle D|\right)$
and $|D\rangle$ ($|A\rangle$) is an electronic state on $D$ ($A$).
$\hat{H}_{osc}=(\hat{a}^{\dagger}\hat{a}+\frac{1}{2})\hbar\omega_{o}+\sum_{q}(\hat{a}_{q}^{\dagger}\hat{a}_{q}+\frac{1}{2})\hbar\omega_{q}$
is the oscillator Hamiltonian for the odorant and environment, and
$\hat{H}_{e-osc,X}=\gamma_{X}\left(\hat{a}+\hat{a}^{\dagger}\right)+\sum_{q}\gamma_{qX}\left(\hat{a}_{q}+\hat{a}_{q}^{\dagger}\right)$
couples the electron to the oscillators. The eigenstates of $\hat{H}_{osc}$
are $|nN\rangle$, where $n$ is the odorant oscillator occupancy
and $N$ corresponds to a set of environment oscillator occupancies
$\left\{ n_{q}\right\} $. The eigenstates of $\hat{H}_{X}$ are $|\Psi_{XnN}\rangle=\exp(u_{X}(\hat{a}-\hat{a}^{\dagger})+\sum_{q}u_{qX}(\hat{a}_{q}-\hat{a}_{q}^{\dagger}))|XnN\rangle$
and have eigenvalues $E_{XnN}=\varepsilon_{X}+(n+\frac{1}{2}-u_{X}^{2})\hbar\omega_{o}+\sum_{q}(n_{q}+\frac{1}{2}-u_{qX}^{2})\hbar\omega_{q}$.
The states $|XnN\rangle$ are products of unperturbed electronic and
oscillator basis states, $u_{X}=\gamma_{X}/\hbar\omega_{o}$ and $u_{qX}=\gamma_{qX}/\hbar\omega_{q}$.
The times $\tau_{T0}$ and $\tau_{T1}$ follow from the standard golden
rule result for the coupling of the system with the electron on $D$
and odorant oscillator in its ground state to that with the electron
on $A$ and odorant oscillator in excited state $|n\rangle$: $1/\tau_{Tn}=(2\pi/\hbar)\sum_{NN'}P_{N}\left|\langle\Psi_{D0N}|\hat{v}|\Psi_{AnN'}\rangle\right|^{2}$
where $P_{N}$ is the probability that the system starts in state
$|\Psi_{D0N}\rangle$. After making standard approximations for an
electron coupled to a bath of phonons\cite{flynn-1970-a,song-1993-a},
and taking the background fluctuations to be of low frequency, we
get the Marcus-type expression

\begin{equation}
\frac{1}{\tau_{Tn}}=\frac{2\pi}{\hbar}t^{2}\frac{\sigma_{n}}{\sqrt{4\pi k_{B}T\lambda}}\exp\left(-\frac{(\epsilon_{n}-\lambda)^{2}}{4k_{B}T\lambda}\right)\label{eq:rate-classical}\end{equation}
where $\sigma_{n}=\exp(-S)S^{n}/n!$, $S=(u_{D}-u_{A})^{2}$ (a Huang-Rhys
factor), $\epsilon_{n}=\varepsilon_{D}-\varepsilon_{A}-n\hbar\omega_{o}$,
$\beta=1/k_{B}T$, $\lambda=\sum_{q}S_{q}\hbar\omega_{q}$ (reorganisation energy), and $S_{q}=(u_{qD}-u_{qA})^{2}$.

\begin{table}
\begin{centering}\begin{tabular}{c|c|c|c|c}
Quantity&
$\hbar\omega_{o}$&
$S$&
$\lambda$&
$|t|$\tabularnewline
\hline
Value&
200 meV&
0.01&
30 meV&
1 meV \tabularnewline
\end{tabular}\par\end{centering}

\caption{\label{cap:values}Estimated values for the physical quantities needed
to compute $\tau_{T0}$ and $\tau_{T1}$. See text for explanation
of their values.}
\end{table}

We now estimate values for the parameters (Table \ref{cap:values}).
The interesting range for $\hbar\omega_{o}$ in olfaction is about
70 meV to 400 meV\cite{turin-2002-a}, so a typical value is 200 meV.
To estimate the Huang-Rhys factor $S$ we introduce a physical mechanism
for the electron-oscillator interaction based on the long-ranged electrostatic
interaction between the electron and odorant atomic partial charges.
The definition $S=(u_{D}-u_{A})^{2}$ is equivalent to $S=\Delta F^{2}/2\hbar M_{o}\omega_{o}^{3}$
where $\Delta F$ is the change in force on the odorant oscillator
as a result of the electronic transition\cite{huang-1950-a,stoneham-2001}.
We treat the oscillator as a dipole with charges $\pm qe$ and compute
the forces on the oscillator when the electron is on $A$ and $D$
(treated as pointlike), giving

\begin{equation}
S=4q^{2}\frac{m_{e}}{M_{o}}\left(\frac{Ry}{\hbar\omega_{o}}\right)^{3}\left(\frac{\hat{R}_{D}\cdot\hat{p}}{(R_{D}/a_{0})^{2}}-\frac{\hat{R}_{A}\cdot\hat{p}}{(R_{A}/a_{0})^{2}}\right)^{2}\label{eq:HR-dipole}
\end{equation}
where $\hat{p}$ is the direction of the dipole, $\vec{R}_{D}$ is the
vector from $D$ to the dipole, $\vec{R}_{A}$ is the vector from $A$ to
the dipole, $m_{e}$ is the electron mass, $Ry$ the Rydberg and $a_{0}$
the Bohr radius. Setting $q=0.2$ (a typical partial atomic charge in a
polar molecule), $m_{e}/M_{o}=1/15000$ (using a representative atomic
mass for light elements), $\hbar\omega_{o}=200$ meV,
$\hat{R}_{D}\cdot\hat{p}=-\hat{R}_{A}\cdot\hat{p}=1$ and
$R_{D}=R_{A}=6\, a_{0}$ gives $S\sim0.01$.

We assume the odorant ($M$) contacts $D$ and $A$ but interacts
with them only weakly with hopping integral $v$. By considering the
resulting admixtures of an $M$ state with energy $\varepsilon_{M}$
with those of $D$ and $A$ we obtain an effective hopping integral
between $D$ and $A$ ( $t=v^{2}/(\varepsilon_{M}-\varepsilon_{A})$).
If $\varepsilon_{M}$ corresponds to a LUMO while $\varepsilon_{D}$
and $\varepsilon_{A}$ correspond to HOMOs then the difference $\varepsilon_{M}-\varepsilon_{A}$
can be as large as 10 eV. The hopping integrals can be estimated for
known molecular structures. Whilst the odorant structure is known,
the donor and acceptor structures interacting with it are unknown,
and we have to make an educated guess. If the bonds between $M$,
and $D$ and $A$ are no stronger than hydrogen bonds, we can put
a rough upper bound on the associated hopping integrals of order 0.1
eV, and hence obtain $t\sim1\,\mathrm{meV}$. Our final conclusions
are not sensitive to this value.

Reorganisation energies are typically of order 1 eV, especially in
hydrated systems, which would result in the elastic channel being
much faster than the inelastic. But much smaller values have been
found, and olfactory receptors are hydrophobic. Experiments on charge
separation in mutant reaction centers of the photosynthetic bacteria
\emph{Rhodobacter capsulatus} show reorganisation energies at room
temperature below 0.03 eV\cite{jia-1993-a}. A generally low value
for odorant receptors would be a result of evolutionary optimization
leading to almost no reorganization during the transition. This requires
that $D$ and $A$ are not too close to the aqueous medium in the
cell to prevent significant coupling to the polarization of the water;
thus we conjecture that $D$ and $A$ must lie well within the lipid
bilayer region (see Fig. \ref{cap:atom chain}). The reorganisation
energy can also be reduced if electronic states on $D$ and $A$ are
extended in space\cite{marcus-1956-a}, so residues with delocalized
electrons may be candidates. (For example, the conserved\cite{fuchs-2001-a}
tryptophan on helix 4 and 3 phenylalanines on helix 3. The surrounding
highly variable residues could modify their redox potentials, producing
different receptors. ) We take a value for the reorganisation energy
of 0.03 eV for the table of values.

Substituting the values in Table \ref{cap:values} into
Eq. \ref{eq:rate-classical} for the case of resonance
($\varepsilon_{D}-\varepsilon_{A}=\hbar\omega_{o}$) we get
$\tau_{T0}\sim 87\,\mathrm{ns}$ and $\tau_{T1}\sim 1.3\,\mathrm{ns}$,
which satisfies the condition $\tau_{T1}\ll\tau_{T0}$, and shows that
the overall time for odor recognition is not limited by the
discrimination process. Increasing the reorganization energy to just
50 meV would make $\tau_{T1}>\tau_{T0}$. Thus, provided the
reorganisation energy can be made not much bigger than $k_{B}T$, we
can obtain a large signal to noise ratio. We note that the theory
remains unaltered if a proton tunnel from $D$ to $A$, but the
parameters $t$, $\varepsilon_{D}$ and $\varepsilon_{A}$ will be
modified.

Our analysis indicates that Turin's model is physically viable provided
the receptor has certain properties (notably, very low reorganization
energy) within ranges known from other biomolecular systems. Our model
shows that the overall charge transfer rate is sufficient to permit
detection on the observed timescales, and the inelastic signal can
be made sufficiently large relative to the elastic signal for there
to be an acceptable signal to noise ratio. Lack of information on
local receptor structure limits what can be verified. Our model illustrates
a more general idea of how molecules can actuate receptors selectively.
Lock and key models rely on docking for discrimination, and mechanical
mechanisms for actuation. Selective docking does have a role in our
class of swipe card models, but the crucial discrimination and non-mechanical
actuation processes are different.

\begin{acknowledgments}
Many entertaining and instructive conversations with Luca Turin and
helpful comments from Rudolph Marcus are gratefully acknowledged.
JB and AH are supported by the EPSRC through the IRC in Nanotechnology.


\end{acknowledgments}

\end{document}